\documentclass[preprint,11pt,3p]{elsarticle}
\usepackage[utf8]{inputenc}
\usepackage{amsmath}
\usepackage{amsfonts}
\usepackage{graphicx}
\usepackage{subcaption}
\usepackage[table]{xcolor}
\usepackage{hyperref}

\journal{XXX}

\begin{document}

\begin{frontmatter}

\title{Non-contact Sensing for Anomaly Detection in Wind Turbine Blades: A focus-SVDD with Complex-Valued Auto-Encoder Approach}

\author[unicamp]{Ga\"{e}tan Frusque\corref{cor1}}
\ead{gaetan.frusque@epfl.ch}
%\ead[url]{http://iagoac.github.io/}

\author[glasgow]{Daniel Mitchell}
\ead{d.mitchell.5@research.gla.ac.uk}

\author[glasgow]{Jamie Blanche}
\ead{Jamie.Blanche@glasgow.ac.uk}

\author[glasgow]{David Flynn}
\ead{David.Flynn@glasgow.ac.uk}

\author[unicamp]{Olga Fink\corref{cor1}}
\ead{olga.fink@epfl.ch}

\address[unicamp]{Laboratory of Intelligent Maintenance and Operations Systems, EPFL, Lausanne}
\address[glasgow]{University of Glasgow, University Avenue, Glasgow, G12 8QQ, UK}

\cortext[cor1]{I am corresponding author}

%\author{Gaetan Frusque, Daniel Mitchell, Jamie Blanche, David Flynn, Olga Fink}
%\date{May 2023}
%Affiliation- University of Glasgow, University Avenue, Glasgow, G12 8QQ

%\begin{document}

%\maketitle

\begin{abstract}
The occurrence of manufacturing defects in wind turbine blade (WTB) production can result in significant increases in operation and maintenance costs of WTBs, reduce capacity factors of wind farms, and occasionally lead to severe and disastrous consequences. Therefore, inspection during the manufacturing process is crucial to ensure consistent fabrication of composite materials. Non-contact sensing techniques, such as Frequency Modulated Continuous Wave (FMCW) radar, are becoming increasingly popular as they offer a full view - cross sectional analysis - of these complex structures during assembly and curing. In this paper, we enhance the quality assurance of WTB manufacturing utilizing FMCW radar as a non-destructive sensing modality. Additionally, a novel anomaly detection pipeline is developed that offers the following advantages:
(1) We use the analytic representation of the Intermediate Frequency signal of the FMCW radar as a feature to disentangle material-specific and round-trip delay information from the received wave.
(2) We propose a novel anomaly detection methodology called focus Support Vector Data Description (focus-SVDD). This methodology involves defining the limit boundaries of the dataset after removing healthy data features, thereby focusing on the  attributes of anomalies.
(3) The proposed method employs a complex-valued autoencoder to remove healthy features and we introduces a new activation function called Exponential Amplitude Decay (EAD).  EAD takes advantage of the Rayleigh distribution, which characterizes an instantaneous amplitude signal. The effectiveness of the proposed method is demonstrated through its  application to collected data, where it shows superior performance compared to other state-of-the-art unsupervised anomaly detection methods. This method is expected to make a significant contribution not only to structural health monitoring but also to the field of deep complex-valued data processing and SVDD application. The code and dataset will be made publicly available.
%The code and dataset are available here \footnote{\href{https://github.com/FrusqueGaetan/FMCWframework}{https://github.com/FrusqueGaetan/FMCWframework} }
\end{abstract}


\end{frontmatter}

\section{Introduction}

    %(Context)
    The wind turbine industry is experiencing consistent expansion within its sector,  with projections indicating a compound annual growth rate of 10.1$\%$ from 2022 to 2027. As a result of this growth, the wind turbine industry is expected to increase its market size from $\$$18.75 billion in 2020 to $\$$26.82 billion by 2027 \cite{GlobalNewswire}. The growth of the wind turbine industry is largely  attributed to  advancements in turbine design, such as the use of larger wind turbines, longer turbine blades (measuring over  100m in length) and the implementation of floating wind turbines, allowing for  positioning further from the shoreline \cite{Mitchell2022}. The increasing size and complexity of wind turbine designs have resulted in a greater need for improvements in quality assurance due to the increased risks and complexity when creating larger designs, especially since manufacturing defects can lead to crack formation, propagation under cyclic stresses and defects further in the operational lifecycle \cite{Galappaththi2013,Schubel2013,Gupta2021}.

    Despite  wind turbine blade composites manufacturing being predominantly a manual process, efforts  have been made to improve the manufacturing  method. Automation is an approach which aims at reducing manufacturing cycle duration's alongside consequent manufacturing unit cost reductions \cite{Desai2022}. Improvements in manufacturing techniques to date have mainly consisted of developing automation in the layup technique, positioning of blade composite to be prepared for sanding, trimming and grinding, as well as reducing the number of personnel working at height, particularly for larger blades   \cite{NREL}. To improve the accuracy in automation, digital twins are being explored as means to couple real-world data into simulations that improve design and assembly processes, and allow for an overview of quality assurance in the manufacturing phase \cite{SOLMAN2022272,LIU2021346}. The intention being that by leveraging historical results and data from SCADA and structural health monitoring techniques to inform current manufacturing at pace that reduces defects occurring in the manufacturing stages \cite{ReinforcedPlatsics,Chen2021}. 
    %Insert space
    %(Motivation/Challenge)
    Inspection during the manufacturing process is essential to ensure that composite materials are fabricated consistently. It can minimize the occurrence  of imperfections, which in turn could lead to complications, resulting in cost overruns  and asset downtime. The manufacturing process of composite materials poses  several challenges, which can vary depending on the technique used.  Consistent hardening of the of composite materials when injecting epoxy resin can be challenging and may result imperfections. Hence, there is a need for sensing methodologies  that can verify the quality of the asset as a high quality material. In this research, we employed  a non-contact method using Frequency Modulated Continuous Wave (FMCW) radar to inspect defects  in epoxy resin injection, which are known to pose challenges. The sensor is low power, lightweight and operates in the K-band.  Initial experiments were conducted on replicated defects during Operation and Maintenance (O$\&$M) \cite{Blanche2020}. This work specifically focuses on quality assurance during the production  of composite wind turbine blades.

    %Insert Space
    %What we Propose
%\begin{itemize}
%            \item(What we propose)
%            \item Short review of complex neural networks in here (complex neural network is used in XYZ but there is not alot of work in it).
%            \item SVDD app \cite{tang2022enhanced} Gearbox, \cite{zhang2022anomaly} structural health monitoring / SVDD progress \cite{azim2023deep} graph subspace \cite{ruff2019deep} deepsad/AE Anomaly detection, \cite{qiu2022abnormal} help (Complex Net \cite{lowe2022complex} object detection, \cite{halimeh2022complex} speech enhencement). AE anomaly detection MIMII \cite{dohi2022mimii}, \cite{takiddin2022deep} cyber attack, \cite{chao2021implicit} emerging fault aircraft engine simulator data.
%    \end{itemize}

There are several methods of inspection which are currently available or in development for structural health monitoring within composites. Montazerian \textit{et. al} conducted a comprehensive review of the various types of sensors that can be embedded within composite materials, including piezoelectric, piezoresistive and optical fibre sensors \cite{Montazerian2019}. However, embedding sensors within a composite being cured can alter the curing process, resulting in delaminations. Hassani \textit{et. al} present a comprehensive review of structural health monitoring techniques for composite structures detailing the types of damage that can be present on a composite structure via non-destructive testing techniques and vibration-based damage detection \cite{Hassani2021}. Listed below are some of the critical requirements for sensing methods that emerged as key limitations in the reviewed papers: affordability, continuous monitoring capability, sensitivity to  low level of damage, ability to detect different damage types,robustness to different ambient loading conditions, resistance to measured noise, and resilience to environmental factors  like weather. Gupta \textit{et. al} provide an in depth review of sensing technologies for non-destructive evaluation of structural composite materials \cite{Gupta2021}. The review paper highlights different sensing mechanisms alongside their working principles, setups, advantages, limitations and usage levels for structural composites. The article also provides an insight into current and future research trends for composites and prognostics and health management (PHM). The current research imperative displays a trend towards non-contact sensing, including methods such as digital image correlation, infrared tomography and FMCW. To sum up, current research in the field highlights the difficulties associated with detecting subsurface defects that are present in composite materials. Detecting potential defects or weaknesses in turbine blades during the manufacturing process is critical  for effective prognostics and health management. As the turbine blades become larger, the risks associated with undetected defects in the early phases also increase. The primary objective of this article is to improve the quality assurance of manufactured blades using FMCW radar as a non-destructive tool. By using this technology, manufacturing centres can  increase production to meet demand while ensuring that  high-quality products with extended overall lifecycles and capabilities are delivered to the market.

% DAN - DF comment actioned

    The FMCW radar is a non-destructive sensing method with the ability to address open challenges in composite structure characterisation via surface and subsurface detection of defects. To date, the sensor has been utilised for several defects within porous dielectric materials including delamination, water ingress, embedded materials and subsurface anomalies. The technology has been applied to sandstones representing civil infrastructure, sandwich and monolithic composite wind turbine blades and tarmac with surface salt dispersions\cite{Blanche2020b, Mitchell2020a,9563264, Tang2023a, Blanche2020Deformation} In addition, the sensor has the ability to detect dynamic load deformations where for a dynamically compressed sandstone, the sensor provided up to 30 seconds of warning in sandstone \cite{Blanche2020a}. 
    %A challenge which this article seeks to overcome includes utilising advanced data analysis techniques to enable for detection or classification of anomalies which may exist within the manufacturing phase, inhibiting the operational lifecycle of an asset. The non-contact approach via the FMCW radar could enable for real time data capture and monitoring throughout the manufacturing process. 
    This article introduces novel framework that combines signal processing and deep learning approaches for detecting anomalies during the manufacturing phase. This work demonstrates the significance of employing FMCW radar for non-invasive real-time data capture and monitoring throughout the entire manufacturing process
    For example, during resin injection, the sensor could be positioned to monitor the flow rate of injection and identify areas which require more resin to assure successful composite bonding; leading to quality assurance.  While composite materials have seen an increase in the wind industry, their usage has also increased due to new use cases within aerospace, hydrogen storage tanks, automotive and civil infrastructure, which will also face similar challenges in future to ensure effective manufacture and operational lifecycle \cite{Advanced_Composite_Book,Wang2019,Fang2019}.

The proposed novel framework makes two methodological contributions that leverages the characteristics of the FMCW radar data. 
Firstly, we introduce an unsupervised anomaly detection approach using support vector data description (SVDD) \cite{tax2004support}.
Although SVDD is not a new  technique for anomaly detection, 
it has been recently applied various studies such as \cite{pan2022rolling} for rolling bearing performance degradation assessment, \cite{tang2022enhanced} for detecting faults in gearbox vibrational signals and \cite{zhang2022anomaly} for structural health monitoring. 
The reason behind its recent popularity is its efficient learning capabilities and the ease of interpretation of the obtained results.
Although there are notable extensions for semi-supervised cases \cite{ruff2019deep}, or data that can be mapped to a graph \cite{azim2023deep}, 
utilizing more advanced strategies such as deep models have not demonstrated better results than the standard approach for unsupervised cases \cite{han2022adbench}. 
Our study is to streamline the SVDD task by focusing on the residuals between the expected healthy behaviour and the observed one, enabling us to focus on anomaly attributes that deviate from the healthy distribution. To achieve this, we  propose a new approach called focus-SVDD that removes healthy data features from the dataset. We propose  utilizing an Autoencoder (AE) to eliminate these features, and also modify the SVDD boundary to address any potential overfitting issues  that may arise from using  this model.

%With that in mind, our proposal in this study is to simplify the SVDD task by eliminating from the data features that we know to be healthy, thus focusing on events that diverge from the healthy distribution. 
%We call our approach focus-SVDD. In order to eliminate healthy features fron the dataset, 
%we propose to use an autoencoder and a modification of SVDD bondary in order to take into account the potential overfitting issues of this model. 

Our second methodological contribution involves exploring various representations of the Intermediate Frequency signal, with a focus on the analytical representation. This representation provides a better separation between the effects of round trip propagation and the frequency-dependent reflectivity of the surface target. However, since the analytical signal is complex-valued, an AE capable of handling this type of data is necessary.
The use of complex-valued network \cite{bassey2021survey} is a recent development in this field, with applications in  remaining useful life prediction of
rotating machinery \cite{russell2022physics}, object discovery \cite{lowe2022complex} and speech enhancement \cite{halimeh2022complex} .
One major challenge faced in this domain is the lack of a standard activation function, such as the Rectified Linear Units (ReLU), used for real-valued autoencoders. In this work, we introduce a novel complex activation function called Exponential Amplitude Decay (EAD) that leverages the Rayleigh distribution of the complex amplitude distribution to achieve better performance.

%-Paper structure
%\section{Literature Review}
%\begin{itemize}
%    \item State we have done reviews on NDE techniques in other papers
%    \item We focus on ML/signal processing techniques for Radar inspection
%    \item Comparitive Analysis table/diagram to justify the route that we have taken.
%\end{itemize}

\section{Materials and Methods}
\subsection{Investigation Procedure: Monolithic Composite}

\begin{figure}[h]
\centering
\hspace*{0.5cm}
\includegraphics[width=4.5in]{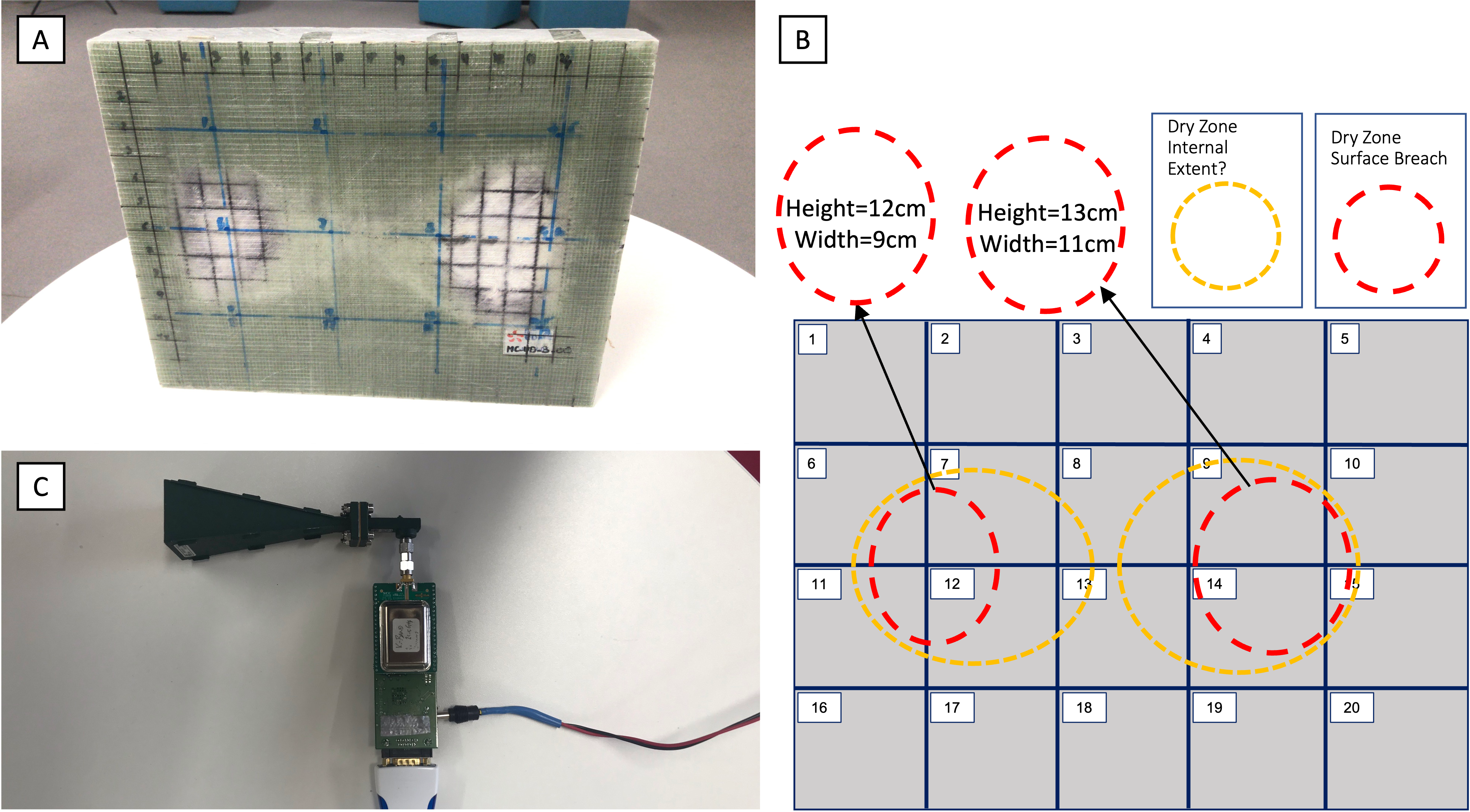}

\caption{ A- Composite with dry adhesive defect, B- Schematic of the monolithic material and C- FMCW radar setup.}
\label{fig:FMCW_Mono_Overview}
\end{figure}

The material under test included a monolithic composite material representing a sample of a wind turbine blade and can be viewed in Fig.~\ref{fig:FMCW_Mono_Overview}A with detailed schematic in Fig.~\ref{fig:FMCW_Mono_Overview}B. The thickness of the material measured 51.4mm \textpm 0.4 at 18 points across the material. The material exhibited a dry adhesive defect typically found during the manufacturing process and is formed during the resin infusion process. This occurs when  insufficient resin penetrates the fibreglass material resulting in an uncured adhesive. The regions of uncured adhesive measured approx. 12 cm in height and 10cm in width. 

\begin{figure}[h]
\centering
\hspace*{0.5cm}
\includegraphics[width=4.5in]{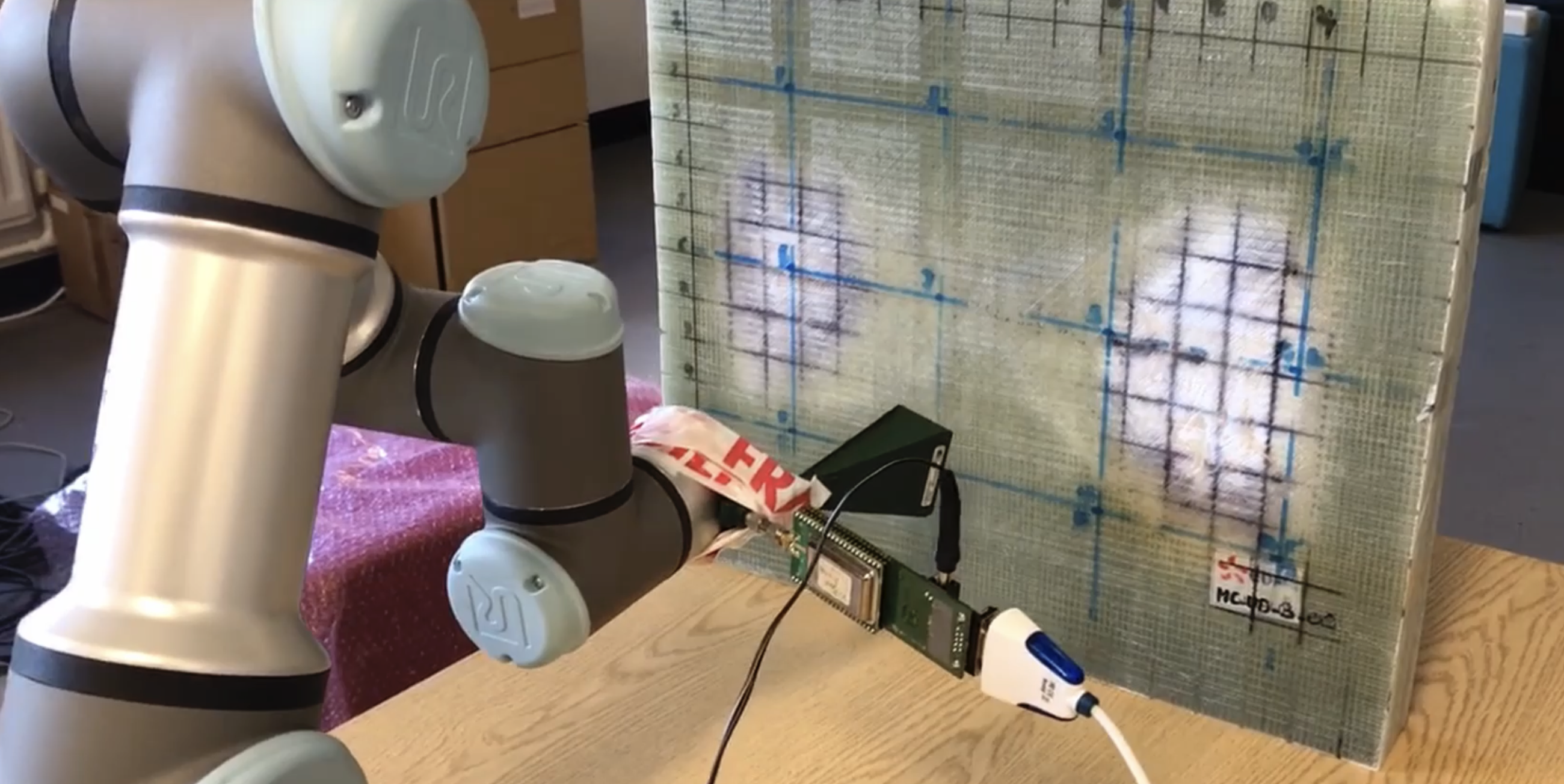}

\caption{The data collection technique using FMCW radar and manipulator arm.}
\label{fig:FMCW_Robot_Arm}
\end{figure}

Fig.~\ref{fig:FMCW_Mono_Overview}C depicts the FMCW radar setup comprising the electronics that generate the electromagnetic wave and the horn antenna that is attached to it. The controller board requires a 12V supply and contains a serial port for data extraction. The sensor weighs approximately 400 grams, consumes low power and can detect surface and subsurface defects within dielectric materials and composites. The radar operated from 24-25.5GHz, bandwidth of 1500MHz with chirp duration of 300 milliseconds and sample rate of 0.2Hz. To ensure an accurate data collection process, we employed  a universal robotic manipulator to automate the process (Fig.~\ref{fig:FMCW_Robot_Arm}.). Several waypoints were positioned in advance of the scan to cover the full area of the composite. We maintained a consistent distance of 10cm from the surface interface. This approach ensured continuous scanning of  different areas of the sample.

\subsection{Data representation}
%Add explaination of the variable x
%normalisaiton explanation
%Add explaination of what we get when we compute the hilbert transform and why it is relevant
%Add a bit mor explaination about the instantaneous amplitude
%Add a small paragraph about fft analysis or not
%Add citation of 1-3 papers
The raw data obtained from the radar sensor is in the time domain and consists of 1501 data points. The signal output from the FMCW sensor is the Intermediate Frequency (IF), which is the difference between the emitted signal and the target interaction with the incident wave. 
According to \cite{piotrowsky2019antenna}, the IF signal, noted here $\mathbf{x} \in \mathbb{R}^{1501}$, can be modelised as a superposition of N weighted cosine function such as: 
\begin{align}\label{hila}
    x(t)= \sum_{n=1}^{N}{a_n{\rm cos}(w_nt+\phi_n)}
\end{align}
where $N$ is the number of radar target or reflective elements. For each target, $w_n$ and $\phi_n$ are proportional to the round trip propagation time, which is the distance between the target and the radar. The factor $a_n$ is the frequency dependant reflectivity and is a function of the target property and frequency sweep range. As a first step, we standardise the IF signal using the average mean and standard deviation of a set of healthy signals. 

To disentangle information about the round trip propagation and  frequency dependant reflectivity, we propose using the analytic representation of the signals. The analytic representation is based on the Hilbert transform $ \mathcal{I}{[\bullet]}$ \cite{piotrowsky2019antenna}, is shown in the following equation when applied to the IF signal: 
\begin{align}\label{hilb}
    \mathcal{I}{[ \mathbf{x} ]}(t)= \sum_{n=1}^{N}{a_n{\rm sin}(w_nt+\phi_n)},
\end{align} 
The goal of the analytic representation is to propose a natural extension of a real signal in the complex domain by removing the negative frequencies. It can be computed as:
\begin{equation}\label{hil}
    \mathbf{x}^{\rm H}= \mathbf{x} + i \mathcal{I}{[\mathbf{x}]},
\end{equation}
Thus by substituting Equations~\ref{hila} and~\ref{hilb} into equation~\ref{hil}:
\begin{equation}\label{hiln}
    x^{\rm H}(t)=  \sum_{n=1}^{N}{a_n{\rm e}^{i(w_nt+\phi_n)}}.
\end{equation}
For the radar target with the highest amplitude, which is expected to be the monolithic composite material, we expect its frequency dependant reflectivity amplitude to remain unchanged over time due to the cosine function. Additionally, we anticipate that the round trip propagation time to be encoded in the phase of the signal.

In addition, we propose to study the instantaneous amplitude representation of the IF signal. This will put more emphasis on the frequency dependant reflectivity coefficients. By taking the modulus $|\bullet|$ of the analytic representation, we have:
\begin{equation}\label{amp}
    \mathbf{x}^{\rm A}= |\mathbf{x}^{\rm H} |
\end{equation}
%Where $ \mathcal{I}{[\bullet]}$ is the Hilbert transform. (not sure of my explanation) The advantages of this representations is to separate the impact of the delay and gain between the emitted and incident wave into respectively the phase and the amplitude of the analytic representation. Indeed, for the case of the raw IF signal, both information are merged in the amplitude.
%In order to remove the delay interactions, we propose to also study the instantaneous amplitude of the signal, which consist simply in taking the modulus $|\bullet|$ of the analytic representation:
%\begin{equation}\label{amp}
%    \mathbf{x}^{\rm A}= |\mathbf{x}^{\rm H} |
%\end{equation}
\subsection{Anomaly detection with kernel Support Vector Data Description} %AE= Autoencoder
Support Vector Data Description (SVDD) is a method developed for dataset characterisation\cite{tax2004support}. Its purpose to find the boundary of a dataset and has been successfully used for anomaly detection in various applications \cite{tang2022enhanced}, \cite{zhang2022anomaly}. Analogous to the support vector machine model, SVDD also uses a kernel function to provide flexibility. The most widely used kernel function is the Radial Basis Function (RBF)\cite{tax2004support}. The application of the RBF kernel between two vectors $\mathbf{x}_k$ and $\mathbf{x}_{k'}$ can be written as: 
\begin{equation}
    \mathcal{K}^\gamma(\mathbf{x}_k,\mathbf{x}_{k'}) = e^{-\gamma ||\mathbf{x}_k-\mathbf{x}_{k'} ||_F^2}
\end{equation}
where $||\bullet ||_F$ is the Frobenius norm and $\gamma$ is a free parameter that needs to be carefully choosen. In this work, our focus is solely on the RBF-SVDD method due to its efficiency. The RBF-SVDD can be regarded as a constrained kernel density estimation problem with a density limit $ D^\gamma_{ \mathbf{X}_T} $, which characterizes the boundary between healthy and unhealthy samples. This density limit can be expressed as follows:
\begin{equation}\label{limit}
    D^\gamma_{ \mathbf{X}_T} = \sum_{k=1}^{K} \sum_{k'=1}^{K} \alpha_k \alpha_{k'}  \mathcal{K}^\gamma(\mathbf{x}_k,\mathbf{x}_{k'})
\end{equation}
where $\mathbf{X}_T= \{\mathbf{x}_1,\mathbf{x}_2,...,\mathbf{x}_{K} \}$ is the training dataset containing $K$ healthy samples. Here, $\alpha \in \mathbb{R}^{K}$ is a vector that describes the importance of each training vector to characterise the density distribution of the dataset. The vector $\alpha$ is computed by minimising the following optimisation problem:  
\begin{align}\label{optim}
 \underset{\mathbf{\alpha} }{ \rm argmin} &\hspace{0.5cm} D^\gamma_{ \mathbf{X}_T} \\ 
\hspace{0.5cm} {\rm s.t.} \hspace{0.5cm} & \sum_{k=1}^{K} \alpha_k = 1,\nonumber \\
\hspace{0.5cm} & 0 \leq \alpha_k \leq C \hspace{0.5cm} \forall k \in \{1,...,K \} \nonumber.
\end{align}
where $C$ is the SVDD hyperparameter that can take values from the segment [$\frac{1}{K}$,$1$]. 
When $C=\frac{1}{K}$, the  optimisation problem has a straightforward solution with $\alpha_k=\frac{1}{K} \forall k$ due to its constraints. In this case, all vectors are support vectors, which  corresponds to a basic case of kernel density estimation. This approach does not include the data at the border or the outliers in the density limit. In contrast, when $C=1$, the optimisation problem is less constrained, and the support vectors represent  the border of the dataset. 
%and it minimise the values of the density function inside the limit of the dataset.
All training data are then included in the density limit, and the optimisation problem, shown in Equation.~\ref{optim} can be solved using linear programming.
Considering the optimal values of $\alpha$ for a fixed parameter $C$ and the training dataset $\mathbf{X}_T$, one can compute the value of the density $d^\gamma_{ \mathbf{X}_T}(\mathbf{x}) $ for any sample $\mathbf{x}$ using the following formula: 
\begin{equation}\label{density}
    d^\gamma_{ \mathbf{X}_T}(\mathbf{x} ) = \sum_{\mathbf{x}_k \in  \mathbf{X}_T}^{} \alpha_k  \mathcal{K}^\gamma(\mathbf{x}_k,\mathbf{x}) 
\end{equation}
We can use the previously computed density limit $D^\gamma_{\mathbf{X}_T}$ to create a decision function $\mathcal{F}^\gamma(\mathbf{x})$ that detects if the sample $\mathbf{x}$ is part of the healthy training dataset, labelled as 0, or an abnormal sample, labelled as 1, based on its density value. If the density value of $\mathbf{x}$ is below the density limit, it is considered an abnormal sample, and if it is above, it is considered part of the healthy training dataset. We have:
\begin{align}\label{E-des}
\mathcal{F}^\gamma(\mathbf{x}) = \begin{cases}  & 0\text{ if      }  d^\gamma_{ \mathbf{X}_T}(\mathbf{x} )  \geq D^\gamma_{ \mathbf{X}_T}  , \\
 &   1\text{ if      }  d^\gamma_{ \mathbf{X}_T}(\mathbf{x} ) < D^\gamma_{ \mathbf{X}_T}.
  \end{cases}
\end{align}
\subsection{Focus-SVDD for Abnormal Feature Detection} 
\subsubsection{Focus by Removing Information about the Healthy State}
While the SVDD aims to find the distribution of the healthy condition from the training dataset, it can be difficult to assess this distribution under different operating conditions of a system. In this work, we propose not to apply the SVDD directly to a raw dataset. Instead, we suggest applying it to a residual dataset where the new samples,  denoted as $\mathbf{r}$ can be computed as: 
\begin{align}
 \mathbf{r} = \mathbf{x} - \mathcal{H}(\mathbf{x}).
\end{align}
Here, $\mathcal{H}(\bullet)$ is a function that computes the best explanation of $\mathbf{x} $ while assuming  it is part of the healthy dataset. Thus, $\mathbf{r}$ only contains values that indicates a deviation from the healthy state. 
We propose to use an autoencoder trained only with healthy data to estimate the function $\mathcal{H}(\bullet)$ and denote the approximation as ${\rm H}_{\theta}(\bullet)$ where $\theta$ represents the trainable parameters of the autoencoder. To train the autoencoder, we  aim to find the values of $\theta$ that minimise the error between the input and its reconstruction: 
\begin{align}
 \underset{\mathbf{\theta} }{ \rm argmin} \sum_{\mathbf{x}_k \in  \mathbf{X}_T}^{} \mid \mid \mathbf{x}_k -{\rm H}_{\theta}(\mathbf{x}_k)  \mid \mid_F^2,
\end{align}
Notice that autoencoders are widely used for anomaly detection in various fields  \cite{doho2022mimii}, \cite{chao2021implicit}, \cite{le2023attention}. 
%Similarly to here, they are train using healthy samples only and a decision is made according to the norm of its residual. In that case, it is related to consider the residual data follows am isotropic multivariate gaussian distribution. In practive, the residual distribution may be more complex, 
%and we propose to not reduce the residual to its norm only, but to estimate a more complex bondary according to SVDD. Moreover, we expect to have more variation of density values when the residual of an abnormal sample is considered. 
Similar to the SVDD method, the training of this model also involves using only healthy samples, and the decision is based on the residual norm \cite{lee2022automated}. However, this approach  assumes the residual data follows an isotropic multivariate Gaussian distribution, which may not always hold in practical applications.
%due to the complexity of residual distribution. 
To address this issue, we suggest estimating a more complex boundary using SVDD instead of using the norm of the residual. Additionally, we anticipate that there will be greater variation in density values when comparing  raw data to abnormal sample residuals.
\subsubsection{Anomaly Detection with Focus-SVDD}
We denote the residual training dataset $\mathbf{R}_T$ as follows:
\begin{equation}
    \mathbf{R}_T = \{ \mathbf{x}_1-{\rm H}_\theta(\mathbf{x}_1), \mathbf{x}_2-{\rm H}_\theta(\mathbf{x}_2), ... , \mathbf{x}_K-{\rm H}_\theta(\mathbf{x}_K) \}
\end{equation}
The RBF-SVDD is applied to the residual data and solves  the optimisation problem in Equation~\ref{optim} using the density limit $ D^\gamma_{ \mathbf{R}_T} $. The density limit $ D^\gamma_{ \mathbf{R}_T}$ is obtained by computing the value of  Equation~\ref{limit} using the residual training data instead of the raw data. By using the optimal values of $\alpha$, this leads to a new density function $ d^\gamma_{ \mathbf{R}_T}(\bullet) $, which is given by Equation~\ref{density} and computed using the residual data. Therefore, the decision function of focus-SVDD is:
\begin{align}\label{E-des2}
\mathcal{F}^\gamma_r(\mathbf{x}) = \begin{cases}  & 0\text{ if      } d^\gamma_{\mathbf{R}_T }(\mathbf{x}-{\rm H}_{\theta}(\mathbf{x}) ) \geq D^\gamma_{\mathbf{R}_T }  , \\
 &   1\text{ if      } d^\gamma_{\mathbf{R}_T }(\mathbf{x}-{\rm H}_{\theta}(\mathbf{x}) ) < D^\gamma_{\mathbf{R}_T }.
  \end{cases}
\end{align}
\subsubsection{Density Limit Correction}
%However, one of the main issue with using an autoencoder is the difficulty to assess how much it overfit to the training dataset. 
A key issue when using autoencoders is the difficulty in  assessing how much the network overfits to the training dataset.
Furthermore, when applied to a dataset with different operating conditions, the SVDD may lack robustness. In this work, we propose a simple modification of the density limit,  denoted $D_m^\gamma$, using a validation dataset $\mathbf{X}_V$ containing healthy samples only. We  hypothesize that overfitting of the autoencoder, or a new operating condition, may induce a systematic bias in the density values. To correct this bias, we propose the following modification: 
\begin{equation}\label{modif}
     D_m^\gamma = D^\gamma_{\mathbf{R}_T } - m_T  + m_V.
\end{equation}
where $m_T$ is the mean density value of the training dataset $\mathbf{R}_T$ and  $m_V$ is the mean density value of the validation dataset $\mathbf{R}_V$. The modified decision function of the focus-SVDD can be expressed as follows:
\begin{align}\label{E-des3}
\mathcal{F}^\gamma_m(\mathbf{x}) = \begin{cases}  & 0\text{ if      } d^\gamma_{\mathbf{R}_T }(\mathbf{x}-{\rm H}_{\theta}(\mathbf{x}) ) \geq D^\gamma_m  , \\
 &   1\text{ if      } d^\gamma_{\mathbf{R}_T }(\mathbf{x}-{\rm H}_{\theta}(\mathbf{x}) ) < D^\gamma_m.
  \end{cases}
\end{align}

\subsection{Proposed Activation Function for Complex-Valued AE}
\subsubsection{Complex-Valued AE}
In this paper, we propose to use a simple architecture for the autoencoder composed of linear layers and Rectified Linear Units (ReLU) activation functions \cite{glorot2011deep}. Considering a value $x$, we have ReLU$(x)={\rm max}(0,x)$. However, since the analytic representation $\mathbf{x}^H$ is a vector of complex values, the ReLU activation function cannot be used in that case. There exist some extensions of the ReLU activation function for complex valued autoencoder \cite{bassey2021survey}. To account for the polar representation of a complex number $z=|z|e^{i\phi_z}$, where $\phi_z$ denotes the phase value, the most widely used  extension of the ReLU layer is \cite{arjovsky2016unitary}:
\begin{equation}\label{R1}
    {\rm CReLU}(z) = {\rm ReLU}(|z|{\rm cos}(\phi_z)) + i {\rm ReLU}(|z|{\rm sin}(\phi_z)).
\end{equation}
The application of Equation~\ref{R1} involves applying the ReLU activation function to the real and imaginary parts of the complex number $z$. Another common extension involves soft-thresholding the amplitude of the complex number $z$, which can be written as \cite{trabelsi2017deep}:
\begin{equation}
    {\rm modReLU}(z) = {\rm ReLU}(|z| - b)e^{i\phi_z},
\end{equation}
where $b$ is a trainable parameter. However, this second activation function was shown to perform worse than Eq.~\ref{R1} in practice \cite{trabelsi2017deep}. To explain this behavior, we first make the assumption that the output of a complex-valued linear layer node follows a centered complex Gaussian distribution. This assumption can be supported by applying the central limit theorem. Thus, we can know the distribution of the modulus $r=|z|$ which follows a Rayleigh distribution with parameter $\Omega$: 
\begin{align}
    {\rm p}(r) = 2\Omega r e^{- \Omega r^2}
\end{align}
This distribution has the disadvantage of be skewed and leptokurtic, impeding the learning of the trainable parameter $b$. Moreover, the threshold is done starting from low values, which are the thin-tailed part of the distribution, outputting a data distribution that is still skewed and leptokurtic. 
This is a notable difference between the application of the ReLU activation function on real input assumed to be Gaussian or uniform. 
\subsubsection{Exponential Amplitude Decay Activation Function}
In this work, we propose a new activation function called Exponential Amplitude Decay (EAD). This new extension of the ReLU layer for complex number is denoted as: 
\begin{equation}\label{EAD}
    {\rm EAD}(z) = (1 - e^{-b|z|^2})e^{i\phi_z}
\end{equation}
The proposed EAD activation function has several advantages. Firstly, it is a smooth differentiable function. Secondly, it constrains the amplitude values to remain within the unit circle, which is not the case with the other two activation functions where the output can have arbitrary high values. Finally, when trainable parameter $b=\Omega$, the resulting variable $v={\rm P}(r)=1-e^{-br^2}$ follows a uniform distribution from 0 to 1. This can be proven by observing that ${\rm P}(r)$ is the cumulative distribution function of the Rayleigh distribution. Consequently, the amplitudes distribution is non-skewed and thin-tailed, which is  not the case for the ${\rm modReLU}$ activation function.

Finally, one advantage of the EAD activation function is its ability to compute the mean, variance, skewness, and kurtosis statistics of the amplitudes for any value of $b$. The analytical formulas for these statistics are provided in \ref{app}, which provides interpretability to the output of each layer of the network.

\begin{figure}[h]
\centering
\includegraphics[width=0.8\textwidth]{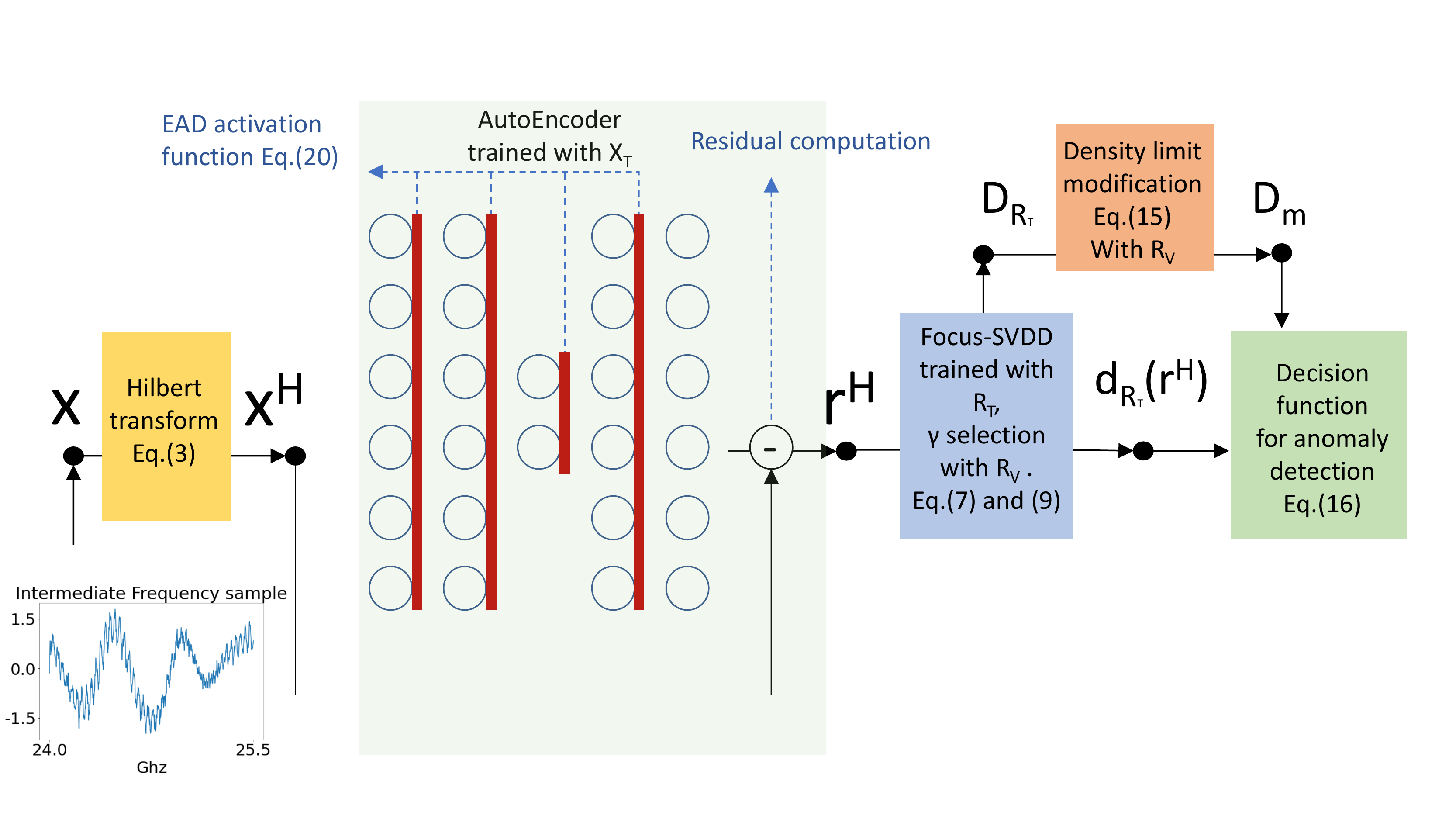}
\caption{Proposed framework for anomaly detection using IF signal of FMCW radars, $\mathbf{X}_{T}$ corresponds to the training dataset, $\mathbf{R}_{T}$ and $\mathbf{R}_{T}$ correspond respectively to the residual training and validation dataset }\label{BigFig}
\end{figure}

The proposed approach is summarized in Figure~\ref{BigFig}

\section{Experiment}
\subsection{Preprocessing and parameter selection}
The dataset was divided into three subsets: the training dataset $\mathbf{X}_T$, the validation dataset $\mathbf{X}_V$, and the test dataset $\mathbf{X}_A$. The training and validation datasets included only samples from the healthy condition, while the test dataset contained both healthy and abnormal samples. A total of 530 healthy samples and 362 abnormal samples were available for testing.
For the experiment, a 5-fold cross-validation was proposed. In this setup, 318 healthy samples were used in $\mathbf{X}_T$, 106 healthy samples were used in $\mathbf{X}_V$, and the remaining 106 healthy samples were used in $\mathbf{X}_A$. The abnormal samples were present only in the test dataset $\mathbf{X}_A$.

To determine the optimal $\gamma$ value for the RBF kernel, we perform a grid search to identify the highest possible value of $\gamma$ at which a fraction $\epsilon$ of the validation dataset is flagged as anomalous. In this study, we have set $C=1$ since we are confident in the labeling of the healthy samples. We have chosen $\epsilon$ to be 0.05 since we have observed that it leads to a more specific decision boundary and yields superior results when using this approach for all methods.
Regarding  the AE, the encoder network consists  of two dense layers of 64 nodes plus one of 32 nodes while the decoder is composed of two dense layers of 64 nodes. We use the ReLU activation function (or one of its complex extensions) in each layer except for the last one.
%the one that maximise the following optimisation problem:
%\begin{align}
%    \underset{\mathbf{\gamma} }{ \rm argmax} &\hspace{0.5cm} \gamma \\ 
%    \hspace{0.5cm} {\rm s.t.} \hspace{0.5cm} &  \frac{1}{K_V}\sum_{x \in \mathbf{X}_V} \mathcal{F}^\gamma(x) < \epsilon.
%\end{align}
%This was proposed to find the maximal value of $\gamma$ over a grid search where a proportion $\epsilon$ of the validation dataset is detected as abnormal. 
%The proposed approach involved conducting a grid search to determine the maximum value of $\gamma$, with a fraction $\epsilon$ of the validation dataset identified as abnormal.
%In this work we fix $C=1$ and $\epsilon=0.05$ since we are rather sure of the labels of the the healthy samples. 

%.......................................................
%We have observed that having a boundary closely aligned with the healthy data yields superior outcomes.

\subsection{Ablation Study 1: focus-SVDD}
We conducted  two ablation studies to evaluate the effectiveness of the one on focus SVDD and the EAD activation function. The first study compares the performance of  focus-SVDD methodology to  standard SVDD. 
Table~\ref{table1a} presents the average F1 scores obtained from five folds when using the real signal $\mathbf{x}$ or its instantaneous amplitude $\mathbf{x}^A$ as inputs for three different decision functions: $\mathcal{F}^\gamma$ (Equation~\ref{E-des}), $\mathcal{F}^\gamma_r$ (Equation~\ref{E-des2}), and $\mathcal{F}^\gamma_m$ (Equation~\ref{E-des3}). 
To remind, $\mathcal{F}^\gamma$ applies SVDD directly to the raw data, while the other two decision functions are instances of focus-SVDD that use either the density limit $D_{\mathbf{R}_T}$ or the modified density limit $D_m$.
We also present the results of applying the standard decision function (Equation~\ref{E-des}) with the modified density limit ($D_m$).
%, which involves replacing the density limit $D_{\mathbf{R}T}$ in Equation~\ref{modif} with the standard density limit $D{\mathbf{X}_T}$.
The results suggest that the performance is comparable when using the standard SVDD decision function $\mathcal{F}^\gamma$, regardless of whether we use the density limit $D_{\mathbf{X}_T}$ or $D_m$. This is because the density values of the validation dataset are not expected to differ significantly from those of the training dataset, and thus $D{\mathbf{X}_T}$ and $D_m$ are similar.
However, there is a significant improvement in performance when using the focus-SVDD methodology and the modified density limit, particularly when considering the real signal as input $\mathbf{x}$. This results in an increase in F1 score of 0.062. Morevoer, the focus-SVDD approach outperforms the standard SVDD anomaly detection method by up to an F1 score of 0.121.

\begin{figure}[h]
\centering
\begin{subfigure}[b]{0.45\textwidth}
        \centering \includegraphics[width=\textwidth]{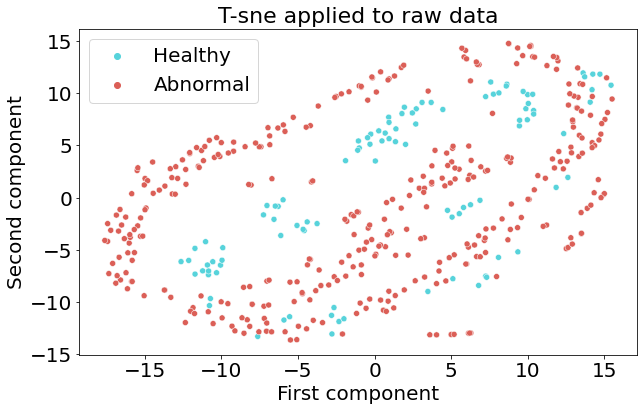}
        \caption{}
    \end{subfigure}
\begin{subfigure}[b]{0.45\textwidth}
        \centering \includegraphics[width=\textwidth]{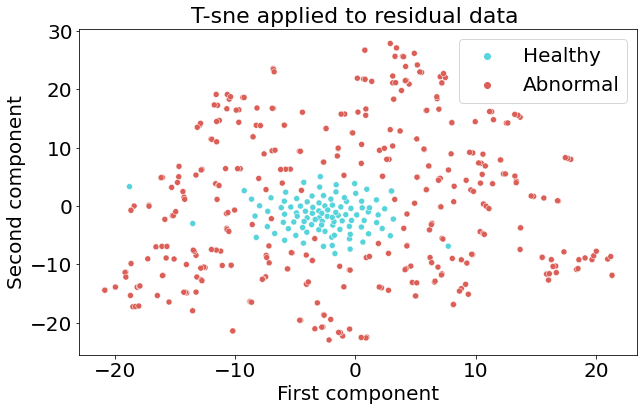}
        \caption{}
    \end{subfigure}
\caption{T-SNE representation with a perplexity of 50 of (a) - the IF test data $\mathbf{X}_A$ (b) - the residual data  $\mathbf{R}_A$ after training an AE on the IF training dataset.}\label{fig_tsne}
\end{figure}

%we provide a t-Distributed Stochastic Neighbor Embedding (t-SNE) \cite{van2008visualizing} with a perplexity of 50 of the IF test data $\mathbf{X}_A$ and the residual data  $\mathbf{R}_A$ after training an AE on the IF training dataset $\mathbf{X}_T$. We can see considering the IF that that the healthy samples belongs to several clusters, and the density limit using SVDD may be difficult to find since it has do adapt to the different operating condition. However, after applying the AE to compute the residual, the dataset become composed of one main cluster. Then the density limit using SVDD only has to describe this cluster.
In Figure~\ref{fig_tsne}, we used t-Distributed Stochastic Neighbor Embedding (t-SNE) \cite{van2008visualizing} with a perplexity of 50 to analyze the IF test data $\mathbf{X}_A$ and the residual data $\mathbf{R}_A$ obtained after training an AE on the IF training dataset $\mathbf{X}_T$. Figure~\ref{fig_tsne} shows that the IF consists of several clusters, each representing a different operating condition. Identifying the density limit using SVDD may be challenging due to the need for adapting to varying clusters. However, after applying the AE to compute the residual, the dataset was predominantly composed of a single main cluster. As a result, using SVDD to determine the density limit only required characterizing this cluster.
\begin{table}
\center
\begin{tabular}{c||c|c|c|c}
\hline 
 & $\mathcal{F}^\gamma$ (Eq.\ref{E-des})& $\mathcal{F}^\gamma$ (Eq.\ref{E-des} with $D_m$) & $\mathcal{F}^\gamma_r$ (Eq.\ref{E-des2}) & $\mathcal{F}^\gamma_m$ (Eq.\ref{E-des3}) \cellcolor{blue!5}  \\ 
\hline 
$\mathbf{x}$ & 0.822 & 0.825 & 0.884 & \cellcolor{blue!5}  \textbf{0.946} \\ 
\hline 
$\mathbf{x}^A$ & 0.795 & 0.795 & 0.916 & \cellcolor{blue!5}  \textbf{0.920} \\ 
\hline 
\end{tabular} 
\caption{Average F1 score over the 5 fold for different signal representation and decision function}\label{table1a}
\end{table}
\subsection{Ablation Study 2: EAD activation function}
In order to characterize the EAD activation function and its advantages over other activation functions, we show in Table~\ref{table1b} the average F1 score over five folds for all three signal representations and the three different complex extensions of the ReLU activation function when using the analytic representation $\mathbf{x}^H$. In the 'Norm' case, the decision is based on the residual norm of the autoencoder, where a proportion $\epsilon$ of the validation dataset is identified as abnormal.
 %Considering the Norm method, we can see the the best performing autoencoder is using the analytical representation $\mathbf{r}^H$ with the EAD activation function which outperform by 0.024 the baseline case using signal $\mathbf{x}$ with a real valued autoencoder. 
The results show that the autoencoder using the EAD activation function and analytical representation $\mathbf{r}^H$ outperforms all other models when using the 'Norm' method, achieving an F1 score  0.024 higher than the baseline case that uses the signal $\mathbf{x}$ with a real-valued autoencoder. Moreover, it outperforms the state-of-the-art complex-valued activation function CReLU by 0.009.
The density limit modification is crucial, as it leads to a significant increase in performance for most cases. Specifically, when using the analytical residual $\mathbf{r}^H$ with CReLU${}$ activation function, we observe an improvement of up to 0.064, except for the case of modReLU${}$, which exhibits a drop in performance of 0.01. Generally, using the $\mathcal{F}^\gamma_r$ decision function leads to decreased performance compared to the 'Norm' decision method. However, using the decision function $\mathcal{F}^\gamma_m$ with modified density limit $D_m$ outperforms the other methods and the best F1 score is obtained by combining the $\mathcal{F}^\gamma_m$ decision function with the EAD activation function, achieving a score of 0.958. 
\begin{table}
\center
\begin{tabular}{c||c|c|c}
\hline 
 & Norm & $\mathcal{F}^\gamma_r$ (Eq.\ref{E-des2}) & $\mathcal{F}^\gamma_m$ (Eq.\ref{E-des3}) \cellcolor{blue!5}   \\ 
\hline 
$\mathbf{x}$ & 0.916 & 0.884  &\textbf{ 0.946} \cellcolor{blue!5} \\ 
\hline 
$\mathbf{x}^{\rm A}$ & 0.841 &  0.916 & \textbf{0.920}   \cellcolor{blue!5}\\ 
\hline 
$\mathbf{x}^{\rm H}$  (CReLU${}$) &  0.931 & 0.878 & \textbf{0.941}  \cellcolor{blue!5}\\ 
\hline 
$\mathbf{x}^{\rm H}$  (modReLU${}$) &  0.884 &  \textbf{0.905} & 0.895  \cellcolor{blue!5} \\ 
\hline 
$\mathbf{x}^{\rm H}$ (EAD)  \cellcolor{blue!5} & 0.940  \cellcolor{blue!5} & 0.928  \cellcolor{blue!5} & \textbf{0.958 } \cellcolor{blue!10}\\ 
\hline 
\end{tabular} 
\caption{Average F1 score over the 5 folds using different activation function and decision function}\label{table1b}
\end{table}

\begin{figure}[h]
\centering
\begin{subfigure}[b]{0.45\textwidth}
        \centering \includegraphics[width=\textwidth]{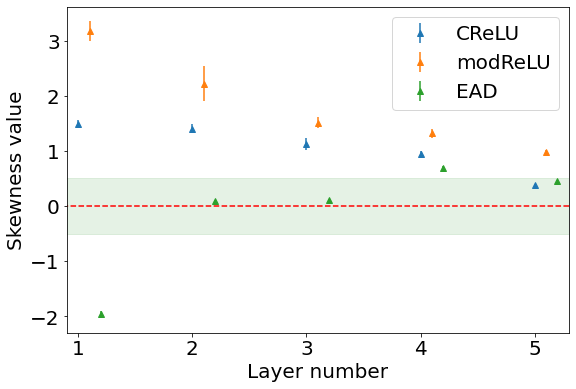}
        \caption{}
    \end{subfigure}
\begin{subfigure}[b]{0.45\textwidth}
        \centering \includegraphics[width=\textwidth]{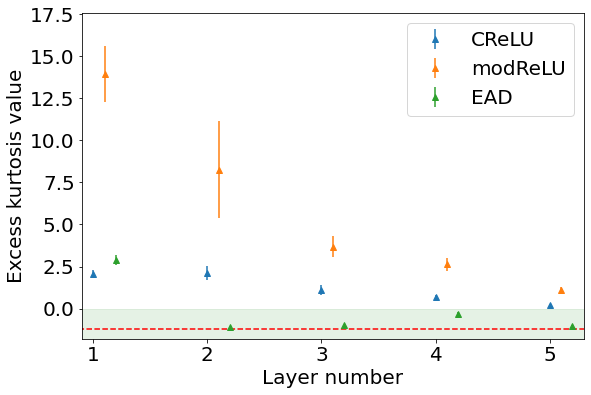}
        \caption{}
    \end{subfigure}
\caption{Mean and standard deviation of (a) Skewness and (b) Kurtosis of the amplitude distribution calculated after each layer of the AE for the three activation functions. In (a) the green area corresponds to low skewness value, in (b) it corresponds to light-tail distribution. The red dotted lines are the statistics of the uniform distribution.}\label{fig_stat}
\end{figure}

In Figure~\ref{fig_stat}, we demonstrate  how the use of our EAD activation function results in the propagation of stable and sub-Gaussian data distribution throughout the network. We display the normalized skewness and excess kurtosis (calculated by subtracting three from the normalized kurtosis) of the amplitude distribution after each layer of the AE for the three activation functions. The green areas represent data distribution with low skewness and platykurtosis  which are indicative of a sub-Gaussian distribution. The red dotted lines depict the statistics of the uniform distribution.
ModRelu and CReLU both produce heavily skewed and leptokurtic amplitudes, indicating a lack of robustness during training. After applying AED, the resulting distribution is in the green area for both statistics in three out of the five layers. For two layers, the statistics are very close to a uniform distribution.
It appears that the proposed activation function is able to improve the network performances by propagating light-tailed amplitude distributions throughout the network.

\subsection{Comparison to other Unsupervised Anomaly Detection Methods}
Finally, we compare the performances of our approach with other state of the art unsupervised anomaly detection method. 
In Table.~\ref{table2}, , we compare the focus-SVDD method with EAD activation function to other unsupervised anomaly detection methods, namely K-Nearest Neighbors using (KNN) \cite{ramaswamy2000efficient}, Isolation Forest (IForest) \cite{liu2008isolation}, DeepSVDD \cite{ruff2018deep}, Local Outlier Factor (LOF) \cite{breunig2000lof}, the AutoEncoder (AE) the Variational AutoEncoder (VAE) \cite{an2015variational}. 
Our implementation of KNN, IForest, DeepSVDD, LOF and VAE are consistent with the APIs of PyOD, which is a comprehensive and scalable Python toolkit for outlier and anomaly detection \cite{zhao2019pyod}. We set the number of neighbors to 5 in KNN and 20 for LOF. We set the number of tree to 100 for the IForest method. Finally, for the DeepSVDD, we use the encoding architecture of the focus-SVDD method with a $\ell_2$ regularisation of 0.001 and for the VAE we use exactly the same architecture as the focus-SVDD methodology. For the VAE, the loss function is modified because a second term is added: the Kullback-Leibler Divergence between the unit Gaussian and the latent distribution, however, only the reconstruction error is used for the anomaly score. 
Since those method are designed  for real data, we use the real signal $\mathbf{x}$ as input for comparison. 
%All those method provide a score, the decision is made based on the score value such as $\epsilon$ sample of the validation dataset are considered as abnormal samples. 
All the previously mentioned methods provide a score. In order to keep consistency with the focus-SVDD, the decision is made based on the score value where a proportion $\epsilon$ of the validation dataset are considered as abnormal samples.
In summary, the performance of each method is evaluated based on F1 score, accuracy, Area Under the Curve (AUC), precision, and recall. The focus-SVDD method outperforms the other models in terms of F1 score, accuracy, AUC, and recall. In contrast, methods such as KNN and Local Outlier Factor (LOF), which use the boundary of the dataset, have a very good AUC score compared to methods that estimate the distribution of the dataset, such as Variational AutoEncoder (VAE) or DeepSVDD. 
%\cite{sokolova2009systematic}

\begin{table}
\center
\begin{tabular}{c||c|c|c|c|c}
\hline 
Method & F1 & Acc. & AUC & Prec. & Recall \\ 
\hline 
\hline 
\textbf{focus-SVDD} \cellcolor{blue!5} & $\mathbf{0.96}$ \cellcolor{blue!5} &$\mathbf{0.93}$ \cellcolor{blue!5} & $\mathbf{0.982}$  \cellcolor{blue!5}& 0.94 \cellcolor{blue!5} & $\mathbf{0.97}$ \cellcolor{blue!5} \\ 
\hline 
KNN & 0.90 & 0.85 & 0.971 & 0.97 & 0.83 \\ 
\hline 
IForest & 0.70 & 0.64 & 0.90 & 0.97 & 0.55 \\ 
\hline 
DeepSVDD & 0.81 & 0.75 & 0.87 & 0.97 & 0.70 \\ 
\hline 
LOF & 0.90 & 0.86 & 0.970 &  $\mathbf{0.98}$ & 0.84 \\ 
\hline 
AE & 0.92 & 0.89 & 0.970 & 0.97 & 0.85 \\ 
\hline 
VAE & 0.87 & 0.81 & 0.95 & 0.97 & 0.78 \\ 
\hline 
\end{tabular} 
\caption{Anomaly detection performances of different methods. (Acc.= Accuracy score, Prec.= Precision score) }\label{table2}
\end{table}

\section{Conclusion}

In this study, we have identified a candidate sensing modality, FMCW radar, for supporting the surface and subsurface analysis of composite - WTB - samples. We utilise a robotic manipulator for the collection of data from the sensing modality. We demonstrate a novel anomaly detection pipeline that utilizes  FMCW radar as a non-destructive sensing modality as to enhance the quality assurance of WTB manufacturing processes. Using composite material samples, the analytic representation of the IF signal of the FMCW radar is utilized to distinguish between  the information related to the round-trip propagation and the frequency dependent reflectivity. We propose a new methodology for anomaly detection, called  focus-SVDD, which involves characterizing the limit boundaries of the residual dataset by removing healthy data features that are mainly related to the operating conditions and are not relevant for the anomaly detection task. Our approach utilizes a complex-valued autoencoder and introduces a novel activation function called EAD, which takes advantage of the Rayleigh distribution to effectively deskew the amplitudes of the output in each layer of the network.
To demonstrate the benefits of utilizing the SVDD on the residual data rather than directly on the raw data, we conduct an ablation study. We also emphasize the significance of modifying the density limit in this scenario as the autoencoder's behavior is susceptible  to overfitting. In addition, we show that our proposed activation function outperforms other complex activation functions and  that utilizing the analytic representation of the IF signal is more advantageous than using the raw IF signal. Finally, we compare our pipeline with a state-of-the-art unsupervised methodology, and the results show superior performance with an F1 score of 0.96, which surpasses the best performing SOTA method by a margin of 0.04. These findings suggest that our approach holds promise for enhancing  the quality assurance of the manufacturing process of WTB and other composite structures.

%DAN
This work  highlights several promising avenues for future research in asset integrity and quality assurance inspections. A trend exists in the robotic deployment of sensors to reduce risks (in both manufacturing and operational phases) and increase accuracy of inspections. A pathway leading to robotic deployment includes testing robotic systems in a controlled environment utilising the non-destructive tool. This enables for data and results from the FMCW radar to be fed into a centralised database to facilitate optimised decision making for quality assurance or damage identification and rectification. Our proposed focus-SVDD approach has the potential to be extended to a semi-supervised approach. where a semi-supervised deep models estimate the boundary of the healthy class. Additionally, the efficacy of the new activation function should be tested in additional applications that utilize complex-valued data. We defer these evaluations for future research.

%A recent trend in SVDD methodologies involves utilizing a trainable neural network instead of a kernel function to estimate the boundary of the healthy class.   
%One potential application for this activation function could be in spectral data processing. We defer these evaluations for future research.

\section{Acknowledgement}
This study was financed by the Swiss Innovation Agency (Innosuisse) under grant number: 47231.1 IP-ENG and the ORCA Hub [EP/R026173/1]. We would like to thanks MicroSense Technologies Ltd (MTL) in the provision of their patented microwave sensing technology (PCT/GB2017/053275) and EDF R\&D division in the provision of wind turbine composite samples.

Researchers using this data set should acknowledge the Smart Systems Group University of Glasgow and cite this paper and the following paper within their references which have created this data library of FMCW analysis of composites \cite{Tang2023a}.

\appendix
\section{Statistics of the EAD activation function.}\label{app}
Considering that the output of a complex-valued linear layer follows a Rician distribution with parameter $\Omega$, it is possible to estimate the statistics of the amplitudes after the application of the EAD activation function (Equation~\ref{EAD}) for any value of the trainable parameter $b$. The statistics $\mu$, $\sigma$, $\nu$, and $\kappa$, corresponding respectively to the mean, standard deviation, normalized skewness, and normalized kurtosis of the variable $u=e^{-br^2}$, when $r$ follows a Rician distribution, have an analytical solution, and are as follows:

\begin{align}
     \mu   & =   {\rm E}[e^{-br^2}], \\
     & = \int_{0}^{}{e^{-br^2}p(r)dr}, \nonumber \\
     & = 2\Omega\int_{0}^{}{r e^{-(\Omega+b)r^2}dr},\nonumber  \\
     & = \frac{\Omega}{\Omega+b}.\nonumber
\end{align}

\begin{align}
\sigma^2 & =   {\rm E}[(e^{-br^2}-\mu)^2], \\
& =   {\rm E}[e^{-2br^2}] -\mu^2, \nonumber \\
    & = \frac{\Omega}{\Omega+2b} - \left(\frac{\Omega}{\Omega+b} \right)^2.\nonumber
\end{align}

\begin{align}
\nu & =   {\rm E}\left[\left(\frac{e^{-br^2}-\mu}{\sigma}\right)^3\right], \\
& = \frac{1}{\sigma^3} ( {\rm E}[e^{-3br^2}] -3\mu\sigma^2 -\mu^3), \nonumber \\
\sigma^3 \nu   & = \frac{\Omega}{\Omega+3b} - 3\left( \frac{\Omega}{\Omega+b}\right) \left(\frac{\Omega}{\Omega+2b} \right)  +2\left(\frac{\Omega}{\Omega+b} \right)^3.\nonumber
\end{align}

\begin{align}
\sigma^4\kappa & =   {\rm E}\left[\left(e^{-br^2}-\mu\right)^4\right], \\
&= \frac{\Omega}{\Omega+4b} - 4\left( \frac{\Omega}{\Omega+b}\right) \left(\frac{\Omega}{\Omega+3b} \right)  
+6\left( \frac{\Omega}{\Omega+b}\right)^2 \left(\frac{\Omega}{\Omega+2b} \right) 
-3\left(\frac{\Omega}{\Omega+b} \right)^4.\nonumber
\end{align}

When $b= \Omega$, we obtain the statistics of an uniform distribution from 0 to 1, i.e. $\mu=\frac{1}{2}$, $\sigma=\frac{1}{12}$, $\nu=0$ and $\kappa=\frac{9}{5}$ ($-\frac{6}{5}$ considering excess kurtosis).

\bibliographystyle{unsrt}
\bibliography{biblio}

\end{document}